# Little Tricky Logic: Misconceptions in the Understanding of LTL

Ben Greenman[a], Sam Saarinen[a], Tim Nelson[a], and Shriram Krishnamurthi[a]

a Brown University, Providence, RI, USA

**Abstract**

**Context** Linear Temporal Logic (LTL) has been used widely in verification. Its importance and popularity have only grown with the revival of temporal logic synthesis, and with new uses of LTL in robotics and planning activities. All these uses demand that the user have a clear understanding of what an LTL specification means.

**Inquiry** Despite the growing use of LTL, no studies have investigated the misconceptions users actually have in understanding LTL formulas. This paper addresses the gap with a first study of LTL misconceptions.

**Approach** We study researchers' and learners' understanding of LTL in four rounds (three written surveys, one talk-aloud) spread across a two-year timeframe. Concretely, we decompose "understanding LTL" into three questions. A person reading a spec needs to understand what it is saying, so we study the mapping from LTL to English. A person writing a spec needs to go in the other direction, so we study English to LTL. However, misconceptions could arise from two sources: a misunderstanding of LTL's syntax or of its underlying semantics. Therefore, we also study the relationship between formulas and specific traces.

**Knowledge** We find several misconceptions that have consequences for learners, tool builders, and designers of new property languages. These findings are already resulting in changes to the Alloy modeling language. We also find that the English to LTL direction was the most common source of errors; unfortunately, this is the critical "authoring" direction in which a subtle mistake can lead to a faulty system. We contribute study instruments that are useful for training learners (whether academic or industrial) who are getting acquainted with LTL, and we provide a code book to assist in the analysis of responses to similar-style questions.

**Grounding** Our findings are grounded in the responses to our survey rounds. Round 1 used Quizius to identify misconceptions among learners in a way that reduces the threat of expert blind spots. Rounds 2 and 3 confirm that both additional learners and researchers (who work in formal methods, robotics, and related fields) make similar errors. Round 4 adds deep support for our misconceptions via talk-aloud surveys.

**Importance** This work provides useful answers to two critical but unexplored questions: in what ways is LTL tricky and what can be done about it? Our survey instruments can serve as a starting point for other studies.

## 1 Introduction

Linear temporal logic (LTL) has long been a standard for writing property specifications in computer-aided verification. The language can express a variety of real-world phenomena while supporting good decision procedures [79]. It is also a small language, thereby presumably making it easy to learn and understand.

In recent years, LTL has been increasingly used for much more than verification. The old dream of temporal-logic synthesis [52, 65] has seen a revival [2, 3, 12]. LTL has been adapted to enable property-based testing of interactive web applications [60]. Even more intriguingly, roboticists have found intimate connections between LTL and robot planning [4, 28, 49], as a result of which numerous robotics systems now use LTL, e.g., [5, 11, 35, 42, 47, 72, 85]. Indeed, it is sufficiently pervasive that there are now even robotics classes that teach LTL to learners who have no prior experience with formal methods [53].

All these efforts are predicated on a central belief: that users of the logic actually understand it. The quality of verification, synthesis, or planning is only as good as the property statement. If a user writes a property but *misunderstands* what it is saying, there are no safeguards: the tool will blindly apply this property and check or generate the requested behavior, whether or not it was the desired one. It is therefore critical to know whether users accurately understand LTL, which is the focus of this paper.

As a case in point, this project was born of a worrisome incident. Two authors attended a research colloquium about using LTL in robot planning. The speaker, a roboticist, began with a brief introduction to LTL. However, this tutorial contained a mistake, but neither the speaker nor the somewhat LTL-aware audience spotted it. This prompted the authors to wonder whether this phenomenon is wider-spread, even amongst people trained in formal methods.

Concretely, this paper focuses on three directions of LTL understanding:

**LTL to English:** Given an LTL formula, can a reader accurately translate it into English? This is similar to what a person does when reading a specification, e.g., when code-reviewing work or studying a paper.

**English to LTL:** Given an English statement, can a reader accurately express it in LTL? This skill is essential for specification and verification.

**Trace satisfaction:** Given an LTL formula and a trace (sequence of states), can a reader accurately label the trace as satisfying or violating? Such questions directly test knowledge of LTL semantics.

We know of virtually no work that examines human factors in this setting (Section 12).

**Outline** This paper begins by explaining the long-term goals of our work and how researchers in related areas have approached similar goals (Section 2). It then presents the design of our multi-year study (Section 3), the formative component of the study (Section 4), and our method for confirming formative findings (Section 5) before shifting focus to the main results and their implications. The paper concludes with related work (Section 12) and a discussion of next steps (Section 13).

**LTL Primer** For a brief overview of the LTL syntax used in this paper, jump ahead to Table 2 on page 5. For a proper introduction, refer to Appendix B.

**Contributions** Our work makes three main contributions:
- We find errors in all three question categories (Sections 6, 7, and 8).
- We provide a code book of misconceptions (Section 5).
- We provide three instruments to test for misconceptions (Appendix A).

These contributions have implications for four classes of LTL users: learners, educators, tool builders, and designers of new property languages. The code book and instruments are of immediate value to learners and educators, whether in academic settings or industrial ones (e.g., [82]). Knowledge about misconceptions can be used by tool builders to create new learning tools or to issue alerts in existing ones. Lastly, our work is of use to designers of logics and is resulting in a change to Alloy.

In short: it may be folk knowledge that LTL is tricky, but *is it really*, *in what ways*, and *what can we do about it*? We believe this paper offers useful initial answers.

## 2 Background on Misconceptions

In educational psychology and other fields, there is an important difference between a *mistake* and a *misconception*. A mistake is simply an error; it could have occurred for any number of reasons. A misconception, in contrast, refers to having the wrong idea about a topic.

Misconceptions usually reveal themselves through mistakes, but not every mistake is a misconception. For instance, if subjects provide the wrong answer in response to a question, there are many possible explanations: they may have a genuine misconception; they may have misunderstood the question; they may have been tired; their hand may have slipped while checking boxes; and so on. In general, we can only discern a misconception by connecting to an intent.

In the education literature, seminal work by Hestenes [37, 38] introduced the idea of a *concept inventory* (CI). A concept inventory is a multiple-choice questionnaire where each question presents one correct answer and several wrong answers. The wrong answers are not chosen arbitrarily, however, but rather are carefully designed to (hopefully) be bijective with specific misconceptions. Thus, when a subject picks a particular wrong answer, there is a very high likelihood they have the corresponding misconception. For instance, if a study of children finds that they often misinterpret the symbol $\times$ to mean addition, a wrong answer for $4 \times 3$ would be 7.

Because a CI maps mistakes to misconceptions without requiring detailed responses, it is lightweight to deploy and effective for recognizing misconceptions with minimal effort. This paper takes concrete steps towards eventually creating a CI for LTL.

**Creating a Concept Inventory** Unfortunately, creating a CI is extremely labor and cost intensive. It often requires several rounds of interviews with subjects using Delphi processes and other resource-heavy methods [1, 23, 25, 33, 36, 75]. In response,

**Table 1** Survey rounds, questions, and paper outline

| | Round 1 | Rounds 2, 3, and 4 |
|---|---|---|
| Trace Sat. | N/A | Section 6 |
| LTL ▹ Eng | Section 4 | Section 7 |
| Eng ▹ LTL | Section 4 | Section 8 |

Saarinen et al. [70] created a system called Quizius in which the subjects themselves generate questions and answer the questions that other subjects have generated. This is a variant of crowdsourcing, known as *classsourcing* or *learnersourcing* [43, 44], in which participants improve an educational instrument while learning from it. Quizius can thereby uncover some class-wide misconceptions with little guidance from experts.

A central question for the Quizius system is how to decide which question to show at any given moment. Questions that are already generating disagreement may produce more of it, exposing more misconceptions; but new questions, on which there is not yet enough data, may produce interesting responses as well. This generates a choice between exploration of new questions and exploitation of existing ones. The choice corresponds to a multi-armed bandit process [7], which is what Quizius uses to decide which questions to present.

The Quizius paper [70] showed (in a different domain: introductory Java programming) that Quizius produces results comparable to intensive expert work at vastly less cost. An additional feature of Quizius is that it helps reduce expert blind spots [56, 57]. However, we do not use Quizius alone; rather, we combine it with judicious use of expert effort to seed the system and to study responses and assemble the instruments described in this paper (i.e., "experts on the outside, Quizius on the inside").

## 3 Research Study Design

Our work spans two dimensions: three kinds of questions and four survey rounds involving different populations. These are the rows and columns of Table 1. There are thus several logical ways to report our findings in a linear order. One possibility is to proceed column-by-column, following the surveys in chronological order. Another is to proceed row-by-row, focusing on the questions.

We feel a hybrid presentation is best. First, we describe Round 1 in full (Section 4) because it was formative for the later rounds. Thereafter we focus on the question types (which are semantically coherent), presenting data from Rounds 2, 3, and 4 combined (Sections 6 to 8), rather than focusing on chronology. Between these two major parts, Section 5 provides context for the latter rounds.

The rest of this section explains the questions, the survey rounds, and their context.

**Table 2** LTL and Electrum syntax

| LTL | Electrum | English |
|---|---|---|
| G(x) | always(x) | x always holds |
| F(x) | eventually(x) | x holds at least once |
| X(x) | after(x)<br>next_state(x) | x holds in the next state |
| x U y | x until y | y eventually holds and x holds until that point |

### 3.1 Study Questions

Our studies focus on two translation questions. These questions relate to practical uses of LTL (syntax in Table 2), namely, the reading and writing of specifications:

1. LTL to English (abbreviated: LTL ▹ Eng) questions present a short formula and ask for a paraphrase. For example, one question and an acceptable answer follow:
   **Q.** Translate to English: not F(not x1)
   **A.** “x1 is always true”
2. English to LTL (abbreviated: Eng ▹ LTL) questions present an English specification and ask for an LTL formula. For example:
   **Q.** Translate to LTL: “x1 holds after one or more steps”
   **A.** X(F(x1))

As we explain in Section 5, we realized after Round 1 (hence the “N/A” in Table 1) the need for a third question type that directly asks about LTL semantics. These questions do not correspond to a task that an LTL user explicitly performs, though it is implicit in translating to LTL:

3. Trace satisfaction (abbreviated: Trace Sat.) questions present two items: an LTL formula and a trace. They ask whether the trace satisfies the formula. All of our traces consist of a sequence of four states, ordered left-to-right from earliest in time to latest, followed by a fifth “lasso” state that repeats forever. The following example represents each state as a set of true variables:
   **Q.** Is the formula X(x or y) satisfied by this trace?
   {x} {x} {} {x,y} {x}+
   **A.** Yes, because x holds in the second state.

### 3.2 Survey Rounds

We collected anonymized data via four survey rounds spread across a two-year span:

- **Round 1** took place in Spring 2020. It used 90 students enrolled in an upper-level, tool-based, applied logic course taught by an author at a private US university.
- **Round 2** took place in Spring 2021 with the same course, institution, and instructor. The course had only 57 students. We attribute the low enrollment to COVID-19.

**Q.** Translate to LTL: If x1 is ever true at some point, x2 must always be true.

**A.** LTL: [ ]
Rationale: [ ]

**Q.** Translate to English: `F(x1) -> G(x2)`

**A.** English: [ ]
Rationale: [ ]

**Figure 1** Round 1 example questions

- **Round 3** took place in Summer 2021 and used 29 anonymous researchers who had prior exposure to LTL.
- **Round 4** took place in Spring 2022 after a third iteration of the applied logic course at the same institution and with the same instructor. We recruited 11 students.

Round 1 used Quizius with expert seed questions to generate preliminary instruments for the LTL ▷ Eng and Eng ▷ LTL questions. Rounds 2 and 3 used Qualtrics and assessed the Round 1 instruments. Round 4 used Qualtrics and Zoom to record participants' talk-aloud reasoning while they took a survey. More details follow, and the artifact includes copies of the surveys.

#### 3.2.1 Round 1 Details

We administered Round 1 as a required assignment. Prior to it, students received two lectures that covered both LTL and the SPIN model checker [39]. One lecture was in-person; the second was remote. Students also modeled the dining philosophers problem in SPIN during one lab session. Unfortunately, students did not do any additional LTL assignments or exams because the first COVID-19 pandemic shutdowns occurred at this time.

Students completed two Quizius quizzes in Round 1, one for each of LTL ▷ Eng and Eng ▷ LTL. They responded to up to ten examples already in the system and had to submit one more question that they deemed interesting to the pool. The assignment instructions promised full credit for all honest attempts, regardless of correctness. Students received credit through an anonymous email address.

To make sure the first few students would have questions to answer, the instructor seeded Quizius with LTL ▷ Eng and Eng ▷ LTL questions. Each direction began with only eight seeds to avoid biasing the question pool toward these expert-generated questions. The seeds were created as pairs; each Eng ▷ LTL seed is a valid translation of one LTL ▷ Eng seed and vice-versa. Figure 1 presents an example seed pair and shows that both question types asked for a translation and a plain-text rationale.

Students did not receive questions as pairs in the manner suggested by Figure 1. If a student happened to receive a matching pair of questions, these questions would have appeared on separate web pages.

#### 3.2.2 Round 2 Details

We administered Round 2 as a required assignment using the Qualtrics survey platform. Prior to the survey, students received four remote lectures that covered LTL and the

**Q.** Is the formula Red satisfied by this trace? (the final state loops)

{G,B} {R,G,B} {R,G,B} {R,G,B} {R,G,B}+

**A.** Answer: Yes / No

What about the trace made you give that answer? [ ]

**Figure 2** Round 2 example trace satisfaction question

Electrum language [51] as implemented in Forge [73]. (Electrum adds LTL to the Alloy modeling tool [22, 41].) Among other topics, the lectures modeled locking algorithms. Students additionally completed one Electrum lab and two Electrum assignments.

For Round 2, we curated the most interesting questions from Round 1 and designed a survey to take no more than 90 minutes of students' time, excluding breaks. (The median completion time was ultimately 93 minutes including time when the browser tab was idle.) The survey had 19 questions, divided as follows:

- 5 LTL ▹ Eng questions,
- 5 Eng ▹ LTL questions, and
- 9 trace satisfaction questions.

After each part, the survey gave students the opportunity to submit one new question and then encouraged them to take a short break before moving on. As in Round 1, the assignment instructions promised full credit for honest attempts and delivered this credit via anonymous email addresses.

To keep student effort manageable, we made two simplifications. First, we did not ask for rationales (which many students had spent a great deal of time writing in Round 1). We instead asked for their confidence and, optionally, a *near miss*: an incorrect response that another learner might submit. Second, we concretized all questions to ask about a panel with three lights: Red, Green, and Blue.

The Eng ▹ LTL section encouraged students to check the syntax (but not correctness) of their formulas and gave students the option of saying that a specification was inexpressible in LTL. The trace satisfaction questions presented a formula and a trace, both represented as text (Figure 2). They asked for a yes/no judgment about whether the trace satisfied the formula and for a plain-text explanation. The instructions for these questions explained that a trace was encoded as a sequence of sets ordered left to right from earliest in time to latest. In this notation, each set contains one letter for each light that is on at that time: R for Red, G for Green, and B for Blue. The instructions explained that final state was a "lasso" that repeated forever; however, the trace syntax did not show the lasso (there was no + sign in the actual survey).

#### 3.2.3 Round 3 Details

To see whether our results extended beyond the students, we posed similar questions to a more experienced population. We requested research colleagues who work in formal methods and robotics to share a survey amongst researchers in their group. To avoid subjects feeling self-conscious, we did not collect email addresses, IP addresses, or any other personal information (and said so up front). The subjects were not paid

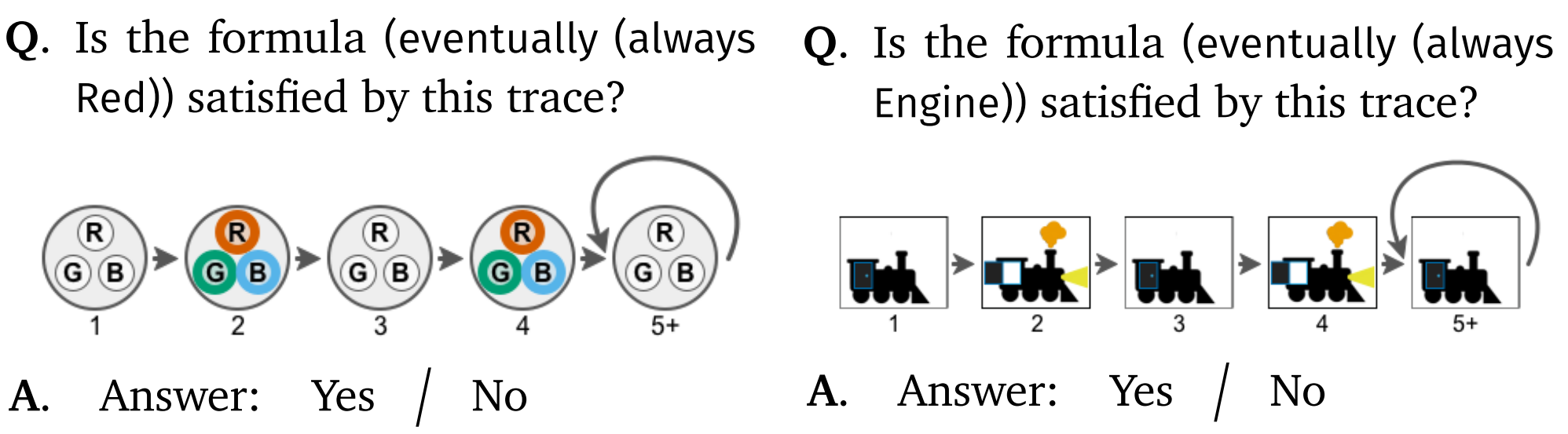

**(a)** Round 3 example

**(b)** Round 4 example

**Figure 3** Rounds 3 and 4 example trace satisfaction questions

for their time. All answers were nevertheless vetted by the authors for "seriousness," and none were excluded on these grounds.

To maximize participation, because we were using the uncompensated time of experts, we wanted to minimize the effort we demanded of them. Therefore, each subject was asked only nine questions from the Round 2 question pool (Qualtrics sampled questions uniformly at random without replacement):

- 4 trace satisfaction questions,
- 3 LTL ▷ Eng questions, and
- 2 Eng ▷ LTL questions — in which we accepted any style of LTL syntax.

The number of questions in each category was inversely proportional to our sense of how much effort each one takes. The final page of this Qualtrics survey collected subjects' research areas and prior exposure to LTL. Most respondents worked (at least partly) in formal methods / verification (N=22, out of 29), almost half worked in AI / machine learning (N=14), three worked in robotics, one worked in programming languages, and one worked in HCI. All respondents had some prior exposure to LTL.

The survey intentionally did not ask subjects to rate their LTL expertise. Doing so at the beginning could have reduced their confidence. Doing so at the end may merely reflect their impression of how they did. Also, respondents may not have shared a frame of reference to answer such a question uniformly.

The Round 3 questions and prompts were the same as in Round 2 with minor changes: we improved the formatting of the LTL ▷ Eng questions, clarified the wording in a few Eng ▷ LTL questions, significantly improved the formatting of the trace satisfaction questions, and removed the rationale for trace satisfaction — to reduce work and thus, hopefully, increase participation. Figure 3a presents an example Round 3 trace. The states are images rather than text, with colored halos around letters to represent lit lights (uncolored = unlit / off), and the transitions are clearly labeled with arrows. The red, green, and blue colors are from a colorblind-friendly palette [84].

#### 3.2.4 Round 4 Details

Unlike the other rounds, we administered Round 4 as a talk-aloud study. Its main purpose was to distinguish mistakes from misconceptions (Section 2) by having subjects explain their reasoning. Its secondary purpose was to test our prior findings in a different domain. We formulated questions in terms of a robotic locomotive with three features — a steam-powered engine, a door, and a headlight — rather than a panel with three lights. Figure 3b presents an example locomotive question; refer to the full survey (in the artifact) for further details about the notation.

We recruited subjects who were enrolled in the Spring 2022 offering of the logic course mentioned above. The material covered in the course did not differ substantially from the previous year (Section 3.2.2). For the study, subjects met with one author for a 25 min Zoom session and received $50 compensation as either an electronic Amazon gift card or a physical gift card to the university bookstore. Subjects completed a Qualtrics survey with 19 questions: 9 trace satisfaction, 5 LTL ▹ Eng, and 5 Eng ▹ LTL. During the study, the author asked subjects to explain their reasoning out loud and provided minor clarifications throughout. For example, if a subject was unsure about the semantics of an LTL operator, the author asked the subject to describe possible semantics, choose one, and continue with that choice.

### 3.3 Additional Details

**On the Use of Students** As Kitchenham et al. [45] point out, the use of students as subjects is not a major threat for studies interested in novice software engineers; after all, "students are the next generation of software professionals." Many of our students had prior work experience and quite a few had job offers by the end of the semester (internship or full-time), making them suitable candidates for our study. All students had some classroom exposure to LTL. In Rounds 2 and 4, students had lab and homework experience modeling systems in Electrum.

**English Language Fluency** Because two of our tasks involve English, the fluency of the subjects matters. Rounds 1, 2, and 4 took place at a university that conducts all classes in English and expects a high degree of fluency; therefore, the findings here may well be a *baseline*. Round 3 participants were either leading researchers or members of their research groups; that is, professionals who read and write papers that use LTL. Again, we did not notice any significant English problems in their written output. Either way, people who read and write formal specifications or scientific papers do so with whatever English language skills they have. If they have a poor understanding of English and this causes misconceptions, this is relevant to our work!

**Two Syntaxes: Classic and Electrum** Although Rounds 1, 2, and 4 took place in instances of the same course, the courses employed different LTL tools and consequently used different syntax for formulas. Table 2 summarizes the syntax involved. Round 1 used the classic LTL syntax with single-letter names for temporal operators: G, F, X, and U. Round 2 switched to Electrum so that students could continue to use Alloy — which was introduced earlier in the course — to work with LTL formulas. Electrum uses English words for the LTL operators: always, eventually, after, and until.

During Round 2, we identified a vernacular misconception regarding Electrum's after operator (discussed in Section 9.1.1). Round 4 and the next iteration of the course therefore used the name next_state for the X operator. Our data suggests that the change is an improvement. (Electrum cannot use the traditional name `next` [65] for `X` because `next` is a reserved word in Alloy.)

**Table 3** Datasets and analysis methods

| Question | Round(s) | Artifact | N | Method |
|---|---|---|---|---|
| Trace Sat. | 2,3,4 | Yes/No choices | 728 | auto |
| Trace Sat. | 2 | Eng text | 513 | manual |
| Trace Sat. | 4 | Eng speech | 99 | manual |
| LTL ▷ Eng | 2,3,4 | Eng text | 427 | manual |
| LTL ▷ Eng | 4 | Eng speech | 55 | manual |
| Eng ▷ LTL | 2,4 | Electrum code | 340 | auto, manual |
| Eng ▷ LTL | 3 | LTL text | 58 | manual |
| Eng ▷ LTL | 4 | Eng speech | 55 | manual |

## 4 Formative Data and Analysis

In Round 1, using Quizius, subjects generated 90 LTL ▷ Eng questions and 87 Eng ▷ LTL questions, which received 901 and 886 answers, respectively. (Not all subjects finished all parts.) These answers were *not* equally spread across the questions because of the multi-armed bandit process in Quizius: questions received a median of six answers, but the highest-ranked questions received 30 or more answers each.

To analyze the data, one researcher studied the high-ranking questions and categorized the incorrect answers. This use of a single "coder" is ameliorated because (a) the coder was working *in conjunction* with the Quizius algorithm; (b) the results were going to be tested in subsequent rounds; and, (c) for generating formal instruments, we intended to and did use multiple coders with a proper inter-coder reliability measure (Section 5). Thus, we felt this was a reasonable process for generating draft instruments.

The data from Round 1 led us to conjecture 15 preliminary LTL misconceptions, listed in the artifact. We intentionally do not present the list here because these preliminaries were mere "stepping stones" that helped us obtain a robust set of misconceptions. For example, one preliminary misconception that did not pan out was the idea that an `X`-wrapped term spans multiple states. Outside of Round 1, very few subjects made mistakes that supported this misconception. Another was about the meaning of the English word "therefore": does it include the current state or not? Rather than test this ambiguity, we chose to avoid it by rewording questions in Round 2. Further details on the preliminaries appear in chapter 7 of Saarinen's dissertation [69].

## 5 Confirming Formative Findings

Armed with preliminary misconceptions, our subsequent rounds set out to confirm that they persist. To do this, we extracted the questions that seemed most productive (in that they generated the most misconceptions), and curated them (e.g., cleaning up their presentation for those that were authored by learners). We used these questions for the subsequent rounds.

After reviewing the Round 1 data, we also realized that we were failing to check how well subjects understood the basic semantics of LTL. This is a subtle, but vital, shift in cognitive setting. *Evaluating* the truth of a formula on a trace and *synthesizing* a formula given an English prompt are related, but not equivalent, skills. This motivates adding trace satisfaction questions.

In fact, traces help in both directions. English answers to LTL ▹ Eng questions may be ambiguous or low-effort. Subjects might simply transliterate (Red U Green) into "Red is on until Green is on," which does not provide much insight. Adding a few trace satisfaction questions therefore lets us check for low-level issues we might have otherwise missed. One such example is the belief that (Red U Green) requires Red to become false when Green becomes true (which we label "ExclusiveU").

### 5.1 Dataset

Table 3 presents a high-level view of the data that we collected in later rounds. Trace satisfaction questions received yes/no answers in all rounds, and additionally received written explanations in Round 2 and spoken explanations in Round 4. LTL ▹ Eng questions received English translations in all rounds and spoken explanations in Round 4. Eng ▹ LTL questions received Electrum code in Rounds 2 and 4, free-form LTL text in Round 3, and spoken explanations in Round 4.

The last column in Table 3 tells us what methods we can use for analyzing the data. Some of these are amenable to automated analysis: e.g., checking the yes/no answers and the correctness of the Electrum formulas (which we checked for *semantic*, not syntactic, equality to the correct answer). The written and spoken English answers require manual analysis. In Round 3, because we did not ask subjects to pre-check their LTL answers (to reduce the time burden), it was simpler to manually classify the responses.

The one remaining entry to explain is the manual analysis of Electrum answers in Round 2. Automated checking lets us decide if a formula is semantically correct. When it is wrong, however, we require human judgment to determine in what way it went wrong. In particular, semantic equality can create bins of equivalence classes, but answers in the same bin may correspond to different misconceptions. This kind of overlap appeared in our dataset; the first "caution" in Section 8 presents an example.

### 5.2 A Code Book for Manual Evaluation

The manual aspects of our analysis use code books to ensure a reproducible mapping from responses to misconceptions. Because we have three question types that require manual analysis (Table 3), we developed three code books. Our coding began with the Eng ▹ LTL responses because we found these were the most difficult to comprehend. Using the misconceptions from Round 1 as a starting point, two authors applied the methods from grounded theory [32] to identify a core set of misconceptions and develop a code book. The coders worked through eight rounds of categorization, independently labeling formulas and then meeting to improve the code book. Each round included at least five and at most 15 formulas. The code book received 18 significant revisions overall, all of which are included in the artifact. After the eight development rounds, the two coders labeled 28 formulas each as a final test for agreement. The Cohen Kappa score [18] on this final test was 0.91, indicating extremely high agreement. In light of the high agreement, one coder labeled the remaining formulas and the second coder merely spot-checked the results.

Next, the two coders scanned the LTL ▹ Eng and trace satisfaction responses for errors that were not covered by the Eng ▹ LTL codes. Although one additional code arose from this scan (ExclusiveU), the coders found that most errors were of a similar nature to the Eng ▹ LTL errors. They therefore adapted the Eng ▹ LTL code book with minor changes to label the LTL ▹ Eng and trace satisfaction responses.

Figure 4 presents a generalized version of the code books. It consists of eleven tags, each of which comes with a description of the erroneous responses to which it applies and one or more examples showing an expected response and an incorrect response. The descriptions use a high-level wording that may be specialized to LTL formulas and English text, such as the formulas and texts in the examples.

The first three tags deal with high-level issues in a particular order: Precedence is for LTL formulas (not English responses) that are correct up to missing parentheses, ReasonableVariant is for LTL formulas that are correct for a slightly different reading of the English question than intended, and Unlabeled is for complex or ambiguous responses. All remaining tags can be studied for any response in any order.

For readers interested in using this code book (either on our instrument questions or on related questions), we offer the following tips and experiences:

- The ImplicitG and ImplicitF responses were of two kinds: they either ignored or added a quantifier. In the LTL ▹ Eng direction, added quantifiers were more common than ignored ones. In the Eng ▹ LTL direction, ignored quantifiers were more common.
- The ExclusiveU tag did not apply to any Eng ▹ LTL responses. Indeed, we recognized the need for this tag only after labeling all LTL responses and moving on to the English responses.
- Neither the BadProp nor the BadStateQuantification tags applied to any LTL ▹ Eng responses. We attribute this to our choice of questions; different questions would likely show a need for both tags.

## 6 Trace Satisfaction

Table 4 tabulates the evidence that we found for various misconceptions among the trace satisfaction questions. The leftmost column lists misconceptions and the

## Abstract Code Book

The following tags describe semantic errors that a learner can make when responding to a survey question. If an incorrect response matches a tag description, we say that the tag *applies* to the response. For Eng ▹ LTL responses, consider the first three tags in order and then consider the remaining tags as a set. For LTL ▹ Eng and trace satisfaction responses, start with the third tag (Unlabeled); if it does not apply, then consider the rest as a set.

1. **Precedence** (Eng ▹ LTL only): Applies to LTL formulas that are correct up to missing parentheses. For other misparenthesized formulas, apply no labels (Unlabeled) instead of guessing about intent.
   - Expected (x and y) => z but subject wrote x and y => z
2. **ReasonableVariant** (Eng ▹ LTL only): Applies to LTL formulas that are correct for an unintended reading of the question. The example here is based on the phrase "Blue will turn on." We intended for Blue to be on now or later, but some subjects (quite reasonably!) specified a flip from off to on. Rajhans et al. [67] call this a *rising edge* issue.
   - Expected F(Blue) but subject wrote (not Blue) and F(Blue)
3. **Unlabeled**: If an answer is convoluted and/or ambiguous, then apply this tag and no others.
4. **BadProp**: Applies to responses that mis-use a logical operator or an atomic symbol.
   - Expected "if x then y" but subject wrote "both x and y hold"
   - Expected G(x) but subject wrote G(y)
5. **BadStateIndex**: Applies to responses that use a correct term at an incorrect state index. Does not apply when a fan-out operator (F, G, U) is missing or included erroneously.
   - Expected "x holds three states from now" but subject wrote "x holds now"
   - Expected x U (y and F(z)) but subject wrote (x U y) and F(z)
6. **BadStateQuantification**: Applies to responses that mis-use or swap a fan-out operator (F, G, U).
   - Expected "x eventually holds" but subject wrote "x always holds"
   - Expected x U y but subject wrote G(x) U y
7. **ExclusiveU**: Applies to responses that assume an until is satisfied only when both the right subterm and the negation of the left subterm hold.
   - Expected x U ((not x) and y) but subject wrote x U y
8. **ImplicitF**: Applies to responses that either ignore or introduce an F quantifier.
   - Expected "whenever x, then y in the next state" but subject wrote "x and y alternate forever"
   - Expected F(x) but subject wrote x
9. **ImplicitG**: Applies to responses that either ignore or introduce a G quantifier.
   - Expected "x always holds" but subject wrote "x holds now"
   - Expected G(x => G(y)) but subject wrote G(x => y)
10. **OtherImplicit**: Applies to underconstrained responses that are not covered by the ImplicitF and ImplicitG tags.
    - Expected "whenever x holds then y also holds" but subject wrote "x holds now and y holds"
    - Expected (not x) U x but subject wrote F(x)
    - Expected x U G(not x) but subject wrote x and F(G(not x))
    - Expected F(x) and G(x => X(G(not x))) but subject wrote F(x and X(G(not x)))
11. **WeakU**: Applies to responses that confuse the U operator with the weak variant W, which does not guarantee that its second subterm eventually holds.
    - Expected "x holds for a finite number of states" but subject wrote "x always holds"
    - Expected x U y but subject wrote F(y) and x U y

**Figure 4** Abstract code book of semantic errors

following columns count evidence from answers in Rounds 2, 3, and 4.[1] The first six rows correspond to misconceptions that we intentionally tested with the trace satisfaction questions. The seventh row shows that four responses gave (unanticipated) evidence of ImplicitG, even though we did not explicitly test for it.

The responses support several misconceptions:

- WeakU (N=29 R2, N=2 R3, N=9 R4), ImplicitF (N=12 R2, N=2 R4), and ExclusiveU (N=9 R2, N=1 R3, N=1 R4) are especially problematic.
- Most of the ImplicitG errors (3 of 4) arose from formulas that constrain a single state (e.g., X(Red)). The other error is from an until formula; it suggests an implicit always on the right side.
- Every BadStateIndex error is due to a mixup about which single state an X-wrapped term refers to.
- The OtherImplicit errors express a desire for F to prevent flickering; specifically, for F(G(Red)) to be satisfied only by traces in which the first Red state is the beginning of an always-Red suffix.

There are far fewer errors among the Round 3 answers. Nevertheless, both WeakU and ExclusiveU may be issues for even these experienced LTL users. Recall also that Round 3 subjects received a random subset of the question pool (Section 3.2.3) rather than every question.

**Cautions** The findings need to be interpreted in the following context:

- A trace satisfaction question pairs a formula with a *specific* trace. It is quite possible that small changes to the trace may affect the responses.
- Round 2 presented traces using text rather than images; refer to Figure 2 for an example. The notation led to identifiable confusion in thirteen responses (23 % R2), but may have misled others in subtle ways.
- Some of the ImplicitF errors may in fact be due to confusion about the start state. Subjects might have assumed that trace matching is allowed to skip a prefix of the trace. However, none of the Round 4 talk alouds contain evidence of this confusion.
- The Round 3 question pool accidentally *omitted* a question that asked whether X constrains two states, the current and next state, rather than the next alone. But given that Rounds 2 and 4 subjects did well on this question (95 % R2, 100 % R4 correct), it seems unlikely that it would have gathered any mistakes.

**Key Takeaway** Subjects had trouble with the basic semantics of F (OtherImplicit), U (WeakU), and unqualified formulas (ImplicitF, ImplicitG).

[1] Note that one answer can increase the counts in multiple rows, provided the response contains more than one mistake. There are no such trace satisfaction responses, but several among Eng ▹ LTL and LTL ▹ Eng data.

## 7 LTL to English

Table 5a shows evidence of misconceptions in translating LTL to English. Overall, 23 % of the Round 2 responses, 15 % of the Round 3 responses, and 24 % of the Round 4 responses were incorrect.

- Half of the Round 2 population and most (9 of 11) of the Round 4 population exhibited the WeakU misconception. Even in Round 3, it caught two people.
- The BadStateIndex errors are of three types, with some overlap: applying the right side of an implication to the next state (N=4 R2, N=1 R4), misinterpreting the scope of an F (N=6 R2, N=1 R3), and misinterpreting the state that an X-wrapped term refers to (N=12 R2, N=2 R3, N=2 R4).
- ImplicitG is common in all populations. In Rounds 3 and 4, an unqualified formula prompted most (N=4 R3, N=4 R4) of the incorrect responses: `R => X(X(X(R)))`. Eight of the ImplicitG mistakes in Round 2 were for the same formula. The others were from a missing G on the right of an until (N=2) and a missing top-level G (N=3).
- The eight OtherImplicit responses are due to one question: `G(Red => X(not Red) and X(X(Red)))`. These responses said that Red blinks forever, which is true only if Red holds in the first or second state.
- One LTL ▷ Eng question tested whether `X(x)` entails `(not x)` in the current state, because several Round 1 responses had that mistake. None of the subjects in Rounds 2, 3, and 4 made this error.

**Cautions** Our findings need some care in interpretation:

- Because the primary output here is English, which is not a formal language, our two-coder method may have mislabeled some written responses. Though in general we found a high level of articulation in the English, there are issues, e.g., some of it was written by subjects who are not native speakers and some of it lacks punctuation.
- The "blinker" errors (OtherImplicit) suggest a kind of confirmation bias among subjects. If the questions had asked for one satisfying and one non-satisfying trace in addition to a translation, subjects might not have made these errors.

**Key Takeaway** In general, subjects did well at this task. The most common error was that subjects expressed a correct constraint at an incorrect time (BadStateIndex).

## 8 English to LTL

Table 5b shows the evidence of misconceptions in translating English to LTL. This corresponds to the direction of authoring LTL based on requirements. The error rates are much higher than for the previous questions: 47 % of the Round 2 responses, 28 % of the Round 3 responses, and 47 % of the Round 4 responses were incorrect. We note that the mistakes are spread across most of the tags.

**Table 4** Trace satisfaction errors

| Misconception | R2 | R3 | R4 |
|---|---|---|---|
| BadProp | 1 | - | - |
| BadStateIndex | 7 | 1 | - |
| ImplicitF | 12 | - | 2 |
| ExclusiveU | 9 | 1 | 1 |
| OtherImplicit | 3 | - | 1 |
| WeakU | 29 | 2 | 9 |
| ImplicitG | 3 | - | 1 |

**Table 5** LTL to English and English to LTL errors

**(a)** LTL to English errors

| Misconception | R2 | R3 | R4 |
|---|---|---|---|
| BadProp | - | - | - |
| BadStateIndex | 18 | 3 | 3 |
| BadStateQuantification | - | 1 | - |
| ExclusiveU | 15 | 1 | 5 |
| ImplicitF | 5 | - | - |
| ImplicitG | 13 | 5 | 4 |
| OtherImplicit | 6 | 1 | - |
| WeakU | 26 | 2 | 6 |
| Unlabeled | 1 | - | - |

**(b)** English to LTL errors

| Misconception | R2 | R3 | R4 |
|---|---|---|---|
| BadProp | 9 | 2 | 6 |
| BadStateIndex | 11 | 3 | 3 |
| BadStateQuantification | 22 | 1 | 2 |
| ExclusiveU | - | - | - |
| ImplicitF | 10 | - | 3 |
| ImplicitG | 47 | 7 | 10 |
| OtherImplicit | 30 | 4 | 8 |
| WeakU | 2 | - | - |
| Unlabeled | 3 | 1 | 1 |
| Precedence | 2 | - | - |
| ReasonableVariant | 18 | 1 | - |

- ImplicitG is the most common error. In all rounds, every question received at least one response that was missing either a toplevel G or one around a subterm.
- OtherImplicit arose in at least two forms. In Round 3, the OtherImplicit all assumed an "eager" semantics for F. Section 9.2.1 discusses this point further. In Rounds 2 and 4, some responses assumed that variables did not change unless specified. Almost all of these were misuses of the F operator. Because variables represented two different things in these rounds — namely, lights on a panel in Round 2 and features on a locomotive in Round 4 — this may be a general issue.
- BadStateQuantification formulas typically contained an extra operator; they rarely swapped one operator for another. A common extra-operator mistake was to write the following formula, which is unsatisfiable: G(Red) U (not Red).
- As mentioned in Section 5, ReasonableVariant applied in cases where subjects used a different interpretation of the question than what we had in mind. We reworded questions between Round 2 and Round 3 in an effort to clarify, which may explain the drop in the presence of this tag.

- BadStateIndex often arose in connection with a binary logical operator. Subjects assumed that the right sides of conjunctions (and) and implications (=>) applied to the *next* state, rather than to the *current* one.
- Although there are few ExclusiveU and WeakU errors, the responses gave us little confidence that subjects can properly use the U operator. There were many misapplications that fell under the BadStateQuantification and BadStateIndex codes.
- LTL has limits on its expressive power [77]; not every property in English can be translated into it. Subjects were aware of this fact, and in Round 1 we saw 4 instances of subjects responding to English sentences with the claim that they were not expressible in LTL. All these responses were incorrect.

  Because the boundaries of LTL expressiveness were not a focus of our study, we presented only English sentences that were LTL-expressible. We did, however, carry over two of the Round 1 questions that prompted "inexpressible" responses. Three such responses arose in Round 2. None arose in Round 3 or Round 4.

**Cautions** Our results carry the following caveats:

- We focused on coding incorrect responses. However, there could be misconceptions lurking in correct responses also. For instance, consider ((x until y) and F(y)) and the same formula without the unnecessary term F(y). The two are semantically equivalent, but the former suggests a WeakU misconception. Through spot-checks we found some evidence for misconceptions lurking in correct answers, but we did not conduct a systematic study.
- Round 2 subjects were accustomed to writing LTL in the context of Electrum, but were asked to fill out the survey checking only for syntactic validity. Access to Electrum's semantic checks may have reduced errors. On the other hand, some subjects may have used their preferred tooling despite the instructions (as we have seen in other settings where we have deployed similar instruments).

**Key Takeaway** By contrast to the other question types, the Eng ▹ LTL direction was fraught with errors and provides evidence for a large number of misconceptions. Unfortunately, this task is perhaps the most important of the three. A user has to write correct LTL in order to apply the logic.

## 9 Implications for Tool Builders and Language Designers

Our findings motivate two concrete suggestions for tool builders and two general suggestions for the designers of LTL-based languages.

### 9.1 Implications For Tool Builders

Tool builders should be aware of the two user-facing issues reported below. The first is about a vernacular misconceptions. The second is about significant whitespace.

#### 9.1.1 "Binary After" Misconception

Upon close inspection, some of the incorrect formulas in Round 2 used the unary after operator (Electrum's version of X) as a binary operator. One example follows, with two red rules (␣) to mark important whitespace:

**Q.** Translate to LTL: Whenever the Red light is on, it turns off in the next state and on again in the state after that.

**A.** always (Red ␣ after (not Red) and (not Red) ␣ after Red )

This is a perfectly natural use of the English word "after," which connects two clauses in the same manner as "until" does; indeed, a unary "after" makes no sense in English. But in Electrum, after behaves quite differently. Electrum reads this formula as having *three* and connectives, two of which implicitly appear at the marked whitespace. Thus, the formula is a syntactically-valid but semantically-incorrect answer.

A binary use of after should ideally be a syntax error. If such a change is not possible, we recommend using a different keyword (such as next_state) to avoid the confusion with English. This problem, known as a *vernacular misconception* [20], has also been found in programming languages [62, 63].

#### 9.1.2 "Implicit-And" Restriction

As the example above demonstrates, Electrum implicitly puts an and between terms separated by whitespace. For the maintainers of large formulas, this shorthand is very helpful. One useful and intentional use of implicit-and is the following predicate for a fully-lit traffic light, which asks whether Red, Yellow, and Green are all in the set of on colors:

```
pred allLightsAreOn
{
  Red in Traffic.on
  Yellow in Traffic.on
  Green  in Traffic.on
}
```

In other formulas, however, implicit-and leads to syntactically-valid formulas that are incorrect in subtle ways.

We raised this issue with the Alloy Board and they agreed to make a change. Henceforth, implicit-and will appear only at newline breaks, and not between space-separated or tab-separated terms. With this change, the traffic light formula remains valid and the formula from Section 9.1.1 raises a syntax error.

### 9.2 Implications For Logic Designers

Designers of temporal logics can directly address misconceptions by changing the logic. Our findings suggest two avenues for improvement.

#### 9.2.1 More Control over Sequencing

Four of the LTL responses from experts in Round 3 had OtherImplicit errors. All four suggest a desire for an F operator that eagerly applies to the first state that satisfies part of its subterm. Here is one example:

**Q.** Translate to LTL: The Red light is on in exactly one state, but not necessarily the first state.

**A.** `eventually(Red and after(always(not Red)))`

The answer is incorrect because it allows traces where Red turns on and off several times before finally staying off. However, it would be correct for a variant of the eventually operator that seeks out the first Red state and tests whether the light is off in the future.

Authors of a new LTL language might consider adding such a variant to their toolbox, or perhaps a Prolog-like cut operator [17] to "commit" at the first occurrence of a Red state. Going further, a language might introduce shorthand for the many applications of F that Menghi et al. [54] propose as core movement patterns; a *strict ordered visit* would suffice here.

A recent robotics paper [5] has the same "eager-F" assumption, and we have also seen it in colloquium talks in robotics. However, this assumption does not seem to be limited to the robotics community: two formal methods researchers in Round 3, twelve students in Round 2, and three students in the Round 4 made the same error.

#### 9.2.2 Explicit State Index

The BadStateIndex errors in our data indicate significant confusion about when an LTL term goes into effect. A language might address these errors by providing more control over the state index, or perhaps an explicit representation of the index.

One common form of BadStateIndex error is the "and then" problem: writing (x and y) instead of (x and X(y)). Subjects in both Round 2 (N=2) and Round 3 (N=2) had this specific problem; others (N=9 R2, N=1 R3, N=3 R4) made a similar mistake with implication. PSL includes a non-overlapping suffix-implication operator with exactly this semantics [27]. Our work provides support for this operator and suggests the need for a suffix-and as well.

A second kind of state index error involved the use of F to connect two toplevel terms (N=4 R2):

**Q.** Translate to LTL: The Red light is on in exactly one state, but not necessarily the first state.

**A.**
```
((not Red) until Red) and
eventually(Red => after(always(not Red)))
```

The answer is on the right track with two useful subterms, but fails to connect them properly. The until subterm requires a non-Red prefix (possibly empty) followed by a Red state. The eventually subterm requires a non-Red suffix after some Red state. The problem is that the two subterms can be satisfied by different Red states—with anything in between.

One could easily fix this particular formula by moving the after term into the right conjunct of the until. But, keeping in mind Electrum's line-based idiom for combining formulas (Section 9.1.2), the human authors and maintainers of LTL specifications might benefit from a more direct representation, perhaps using labels to bind and refer to state indices.

## 10 Implications for Educators

Both our instruments (Appendix A) and our inventory of misconceptions (Figure 4) are each independently of use. Educators may wish to apply these in a traditional classroom setting or as part of a workplace training program. It is worth noting that we cannot guarantee that incorrect answers are bijective with misconceptions, so some further interpretation of wrong answers is necessary.

Once we have found misconceptions, how can they be addressed? A natural tendency is to assume that learners can just be "taught right." While that may work in some cases — e.g., by incorporating these findings into how LTL is introduced — research suggests that that may not suffice.

There is an extensive literature on understanding misconceptions. The US National Research Council [20] refines them into five categories; some, like preconceived notions, are very unlikely to apply here, while others seem especially relevant: vernacular misconceptions are those caused by using words with other natural language meanings, and conceptual misunderstandings are those that arise when instruction permits learners to create faulty models.

Many techniques have been proposed to help learners overcome misconceptions. Researchers have found success using *concept maps* [58]. The work of Posner et al. [66] presents a theory of conceptual change, at the heart of which is the *refutation text* [83]. Studies show that these work in some cases or domains, but sometimes not in others. In general, ideas from physical science education do not necessarily carry over directly to LTL because learners form conceptual understandings of the physical world due to their daily interaction with it (e.g., gravitation), which seems unlikely for most dimensions of understanding LTL. In general, however, the literature makes clear [13, 50, 71] that direct instruction alone is unlikely to overcome misconceptions; activities that are more learner-centric are much more likely to be effective.

## 11 Threats to Validity

**Internal Validity** Coding inherently contains various biases. Our high inter-coder reliability score only indicates that the coders have aligned their biases, not that they have eliminated them. Nevertheless, we believe the codes we have arrived at are reasonable. The artifact lets others review our coding.

Quizius has two threats that can cause it to overlook an interesting question: ordering and timing. If the first few recipients of a difficult question happen to get it correct, then the question receives less attention going forward. And if a question arrives late during the quiz period, it has few opportunities to gather answers.

**External Validity** Rounds 1, 2, and 4 are based on students at the same institution, in the same course, and with the same instructor. Relaxing each of these factors can cause different outcomes. For instance, the instructor's level of comfort with the natural language can be a factor. Even the amount of exposure could help: apparent misconceptions in Round 1 that did not replicate in Rounds 2 and 3 may be because of the limited prep time Round 1 subjects had due to COVID-19. Although Round 3 supports a large number of misconceptions and suggests that our observations generalize beyond the Round 1 population, repeating the process on a much broader population would be valuable and may reveal new misconceptions.

**Ecological Validity** Our survey setting is quite distinct from practical modes of using LTL. Subjects had to answer questions about abstract formulas, absent a concrete use-case, without access to a solver. Studying subjects as they use a tool (perhaps using talk-alouds), while much harder to run in a controlled fashion, would help us identify how these misconceptions are handled in practice.

It is worth noting, however, that using LTL in context (with a tool) will not automatically catch mistakes. Both the system-under-test and the formula could be incorrect in the same way, in which case no counter-example will help the user correct their misconception. For that reason, studies of our kind are still useful and provide concrete targets for more ecologically valid studies.

**Construct Validity** Our work suffers from some construct validity as well. Apparent misconceptions could be artifacts of the specific wording or presentation in our instruments (as we have already seen to some extent: Section 5), or could be evidence of some other misconception that we had not identified, such as the confusion about rising edges [67]. The main way to mitigate these concerns is both to perform further validation steps and to tie our results to methods that improve ecological validity. One other construct validity concern would be if our work was misinterpreted as claiming to capture *all* misconceptions, but we make no such claim.

**Conclusion Validity** Our work is intentionally light on "conclusions," for two reasons. First, as the first work of its kind, it is inherently formative. Second, our main product is *instruments* that others can use; it is those uses that might arrive at conclusions, whose validity the users would have to justify. Our main conclusion here is that there are misconceptions in the understanding of LTL, and that we have instruments that can identify some of them with reasonable confidence. The former seems inherently likely: it is hard to imagine, given all the misconceptions about other formal artifacts, that LTL is a solitary exception. For the latter, we believe our multiple rounds and forms of validation lend some credibility to our instruments.

## 12 Related Work

Despite the long and distinguished history of literature on temporal logic (e.g., [16, 48, 52, 61, 64, 65, 80]), to our knowledge, there are no in-depth user studies on patterns of misconception. The only exceptions we know are two recent studies: one that compares LTL to two similar logics [21], and one that compares three temporal specification languages [15]. Although these studies have some information about how learners fare with temporal formulas, the focus is on comparison *between* languages, not on why learners struggle with LTL specifically.

Of course, it is not news that properties might contain mistakes! Vacuity checking [10, 14, 29, 46, 55], for instance, guards against one kind of error. Other works have addressed other specific specification errors: e.g., [8, 40, 59]. However, all these focus on particular *mistakes*; they do not imply a misunderstanding of the logic itself. Furthermore, even a property that passes the above checks can be *wrong* due to a misunderstanding of the logic. Our focus is thus on understanding, which in turn can generate new checkers.

Several authors have created temporal logic formula templates by examining families of examples. Dwyer et al. [26] identify a taxonomy of specification patterns based on industrial and academic examples. Menghi et al. [54] introduce patterns for robotics missions drawn from hundreds of natural-language specifications. Rajhans et al. [67] identify several template formulas, developed through conversations with industry partners, presented through a graphical tool. Our work is different in its focus on providing fundamental insights into users' misconceptions (often expressed through mistakes), rather than patterns of normal use.

There is a longstanding debate on the relative merits of linear-time (e.g., LTL) versus branching-time (e.g., CTL [16]) logics. This debate is *emphatically not* the topic of this paper; we do not claim to address what is most "intuitive" [78]. If we had begun this process with a different logic, such as CTL, we might well have surfaced other errors not seen here. We stress that it would not suffice to reuse our prompts in a different setting: vital features of the logic might not be exercised and much insight would be missed.

The linear vs. branching debate led to new logics in the early 2000s. IBM introduced Sugar [9], which mixes linear- and branching-time semantics and later grew into PSA [27]. Likewise, Intel released the language ForSpec [6]. Both languages are motivated by the desire to improve the user experience, and both also make central claims about usability: e.g., Sugar claims that hardware engineers can "easily and intuitively specify their designs". However, these claims are asserted without rigorous justification, and neither is accompanied by any catalog of remaining difficulties of the form that we have found.[2]

[2] From personal communication with Moshe Vardi, we learned that many design choices in ForSpec, such as using English words instead of mathematical symbols, were indeed based on extensive (but informal) discussions with different types of users.

In terms of designing new logics, there is a broad literature on misconceptions and design methods in programming languages, such as classic works by Pea [63], Pane and Myers [62], and others. The methods, if not the findings, of these works would be useful to the designers of new logics.

The Wason Selection Task [81] has been used to argue that humans may not reason using the rules of logic (though others have argued the result is contextual [19, 74]). In contrast, our work does not study *reasoning*, only on understanding of the logic itself. However, some results on how context impacts semantics [74] may eventually link the two efforts.

Finally, our English-to-LTL exercises resemble those in Iltis [31], a tool for teaching logic. It is possible that Iltis would provide a good framework for studies like ours, although the tool itself focuses on pedagogy, rather than studies of misconceptions.

## 13 Discussion

**Other Question Types** Our study and results suggest several more types of questions that would yield additional insight. For instance, we did not ask subjects to explicitly create traces corresponding to formulas; in particular, there would be value to asking for both correct and (near-miss) incorrect traces, to probe their understanding of the space described by a formula. One can similarly go from formulas to traces. We could also test their understanding of the (in)equality of pairs of LTL formulas. Other studies [21, 68, 76] contain instruments that would also be useful to adapt.

**Other Logics** Naturally, this style of work can be applied to other property languages as well, such as CTL. Writing correct specifications for complex systems is difficult in any formal language, especially because the requirements and the underlying system may change over time (see discussions in [24, 30]). It is therefore critical to identify specification errors and correct them. Our processes suggest a possible approach, and (for similar logics) our instruments provide a *starting* guide. Comparing logics is important as well (prior work: [15, 21]). One question, suggested by a Programming reviewer, is whether the logic-to-English direction is in general less error-prone than English-to-logic. While perhaps unsurprising if so, affirmative evidence would be very useful to have.

Whatever the process used, we stress the importance of using mechanisms that mitigate expert blind spots such as Quizius. In fact, four questions in the final instruments were generated by subjects in Round 1 (A.2.1, A.3.2, A.3.3,and A.3.5).

**Conclusion** With the use of LTL on the rise, it is important to understand its ergonomics well. Our studies constitute a first, formative step in this direction, and have already had some language design impact. The errors that we found were consistent across two syntaxes, two domains, and a variety of subjects, which suggests that the misconceptions are robust. We also suggest some high-level consequences for the design of new property languages. Given that relatively simple formulas generated our findings, we suspect many other issues lurk in more complicated formulas.

**Data Availability Statement** The data for this paper are on Zenodo [34].

**Acknowledgements** Thanks to Moshe Vardi for valuable conversations about the design of temporal logics. In particular, Shriram thanks Moshe for introducing him to the inestimable beauty of temporal logic. Moshe should probably not be held responsible for Shriram's own misconceptions about LTL!

This work was partially supported by the US National Science Foundation grants DGE-2208731, SHF-2227863, and 2030859 to the CRA for the CIFellows project. This work was also partially supported by RelationalAI.

Thanks to Kathi Fisler, Thomas DelVecchio, Kuang-Chen Lu, Elijah Rivera, and Yanyan Ren for pilot-testing our surveys and suggesting improvements. Thanks to Elijah Rivera and Kuang-Chen Lu for help with the artifact. Thanks to the Programming reviewers for detailed comments.

Last but not least, thanks to all the students in our classes and the researchers who kindly participated in our studies.

# A Instrument

## A.1 Trace Satisfaction

### A.1.1

Is the formula Red satisfied by this trace?

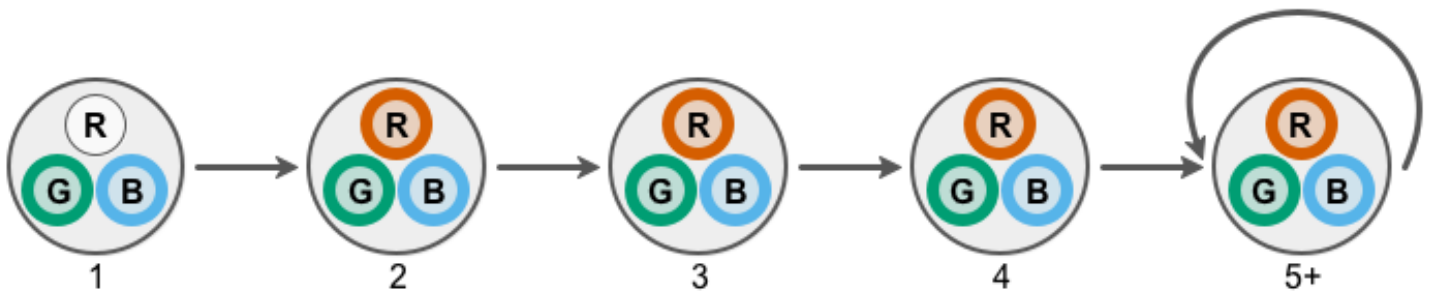

Answer: Yes / No

### A.1.2

Is the formula after(after(after(Red))) satisfied by this trace?

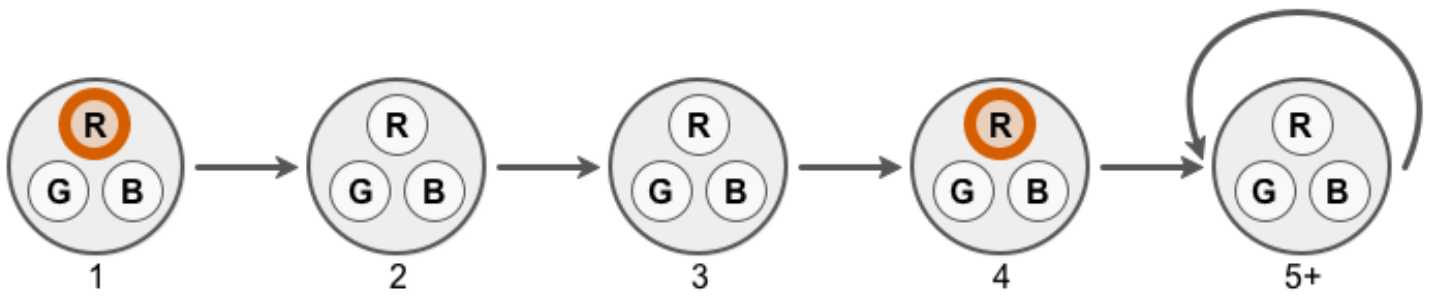

Answer: Yes / No

### A.1.3

Is the formula always(Red => after(after(after(Red)))) satisfied by this trace?

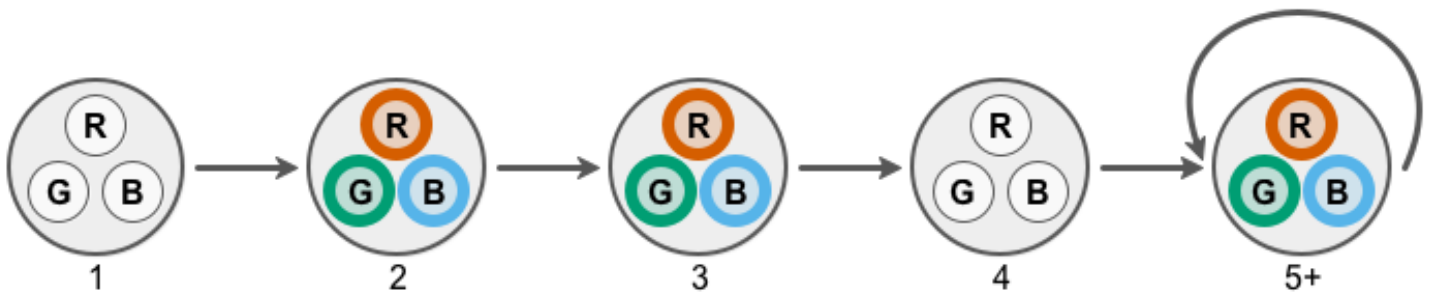

Answer: Yes / No

### A.1.4

Is the formula ((after Red) until (after Green)) satisfied by this trace?

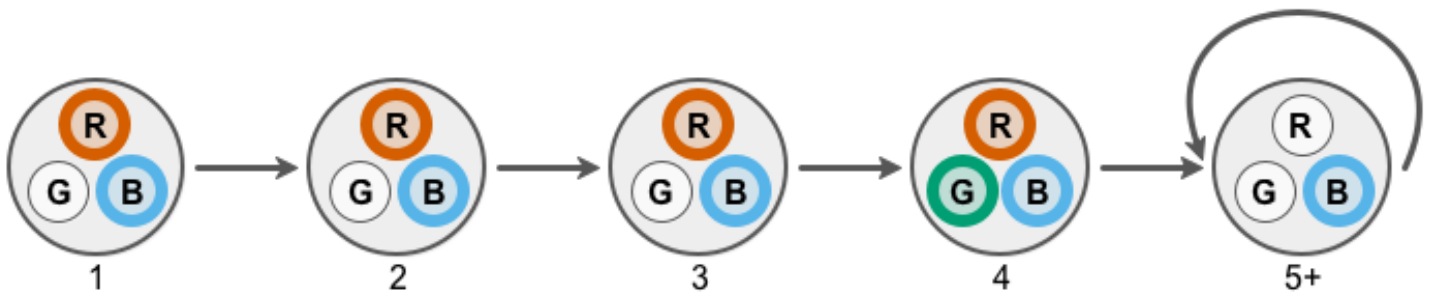

Answer: Yes / No

**A.1.5**
Is the formula ((eventually Red) and (eventually Green)) satisfied by this trace?

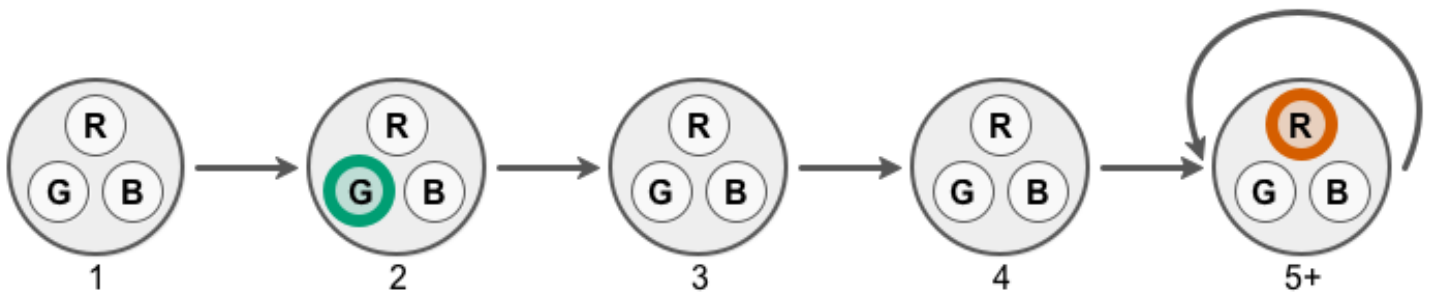

Answer: Yes / No

**A.1.6**
Is the formula after(after(eventually(Red))) satisfied by this trace?

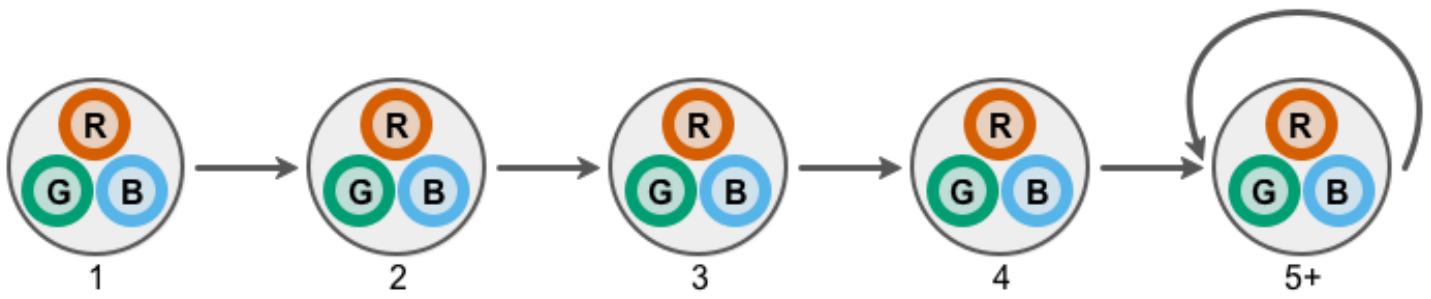

Answer: Yes / No

**A.1.7**
Is the formula (Red until Blue) satisfied by this trace?

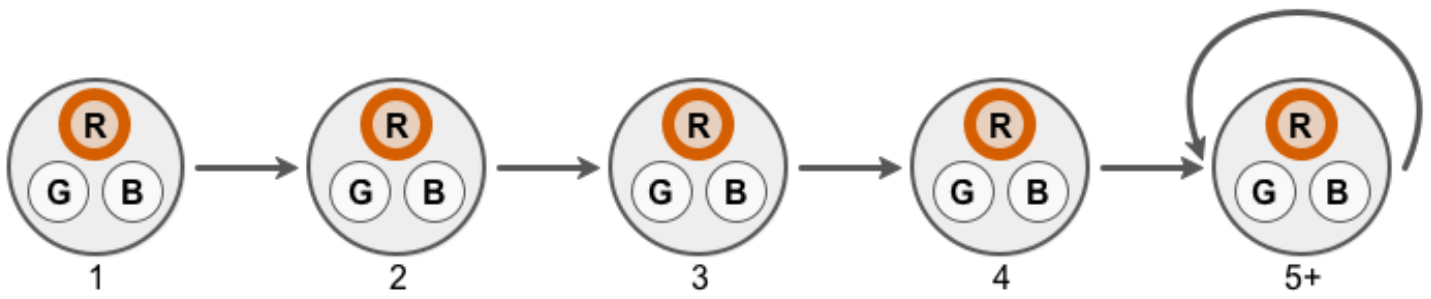

Answer: Yes / No

**A.1.8**
Is the formula eventually(always(Red)) satisfied by this trace?

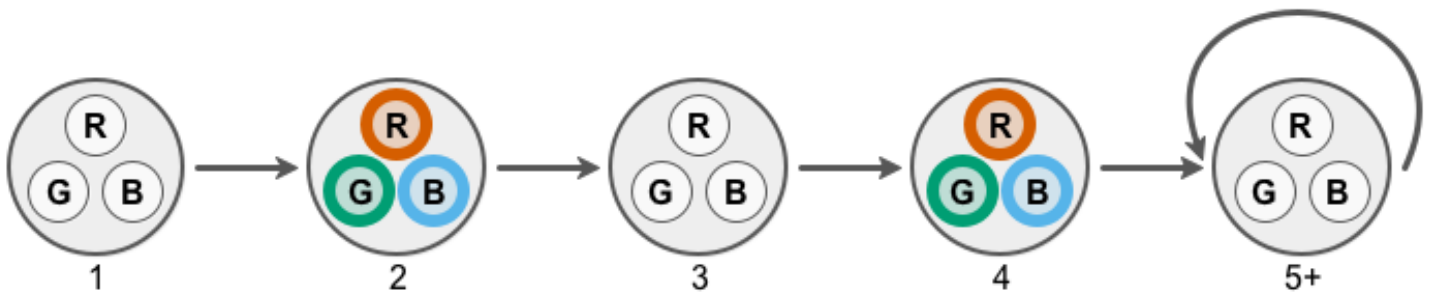

Answer: Yes / No

**A.1.9**
Is the formula always(Red => Green) satisfied by this trace?

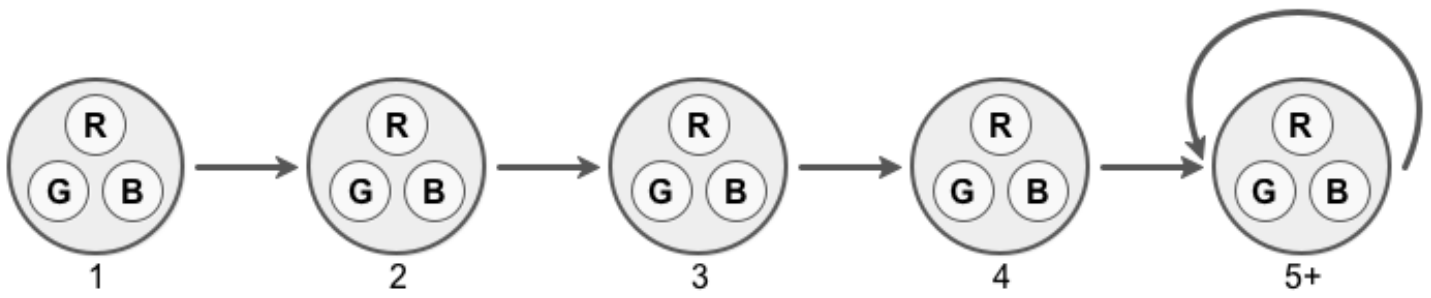

Answer: Yes / No

## A.2 LTL to English

### A.2.1

Translate to English: `Red => after(after(after(Red)))`

Answer: ______________________________

### A.2.2

Translate to English: `after(after(eventually(after(Red))))`

Answer: ______________________________

### A.2.3

Translate to English: `((eventually Red) => (always Blue))`

Answer: ______________________________

### A.2.4

Translate to English: `((Red until Blue) and always(Red))`

Answer: ______________________________

### A.2.5

Translate to English: `always(Red => (after(not Red) and after(after(Red))))`

Answer: ______________________________

## A.3 English to LTL

### A.3.1

Translate to LTL: Whenever the Red light is on, it is off in the next state and on again in the state after that.

Answer: ________________________________________

### A.3.2

Translate to LTL: The Red light is on in exactly one state, but not necessarily the first state.

Answer: ________________________________________

### A.3.3

Translate to LTL: The Red light cannot stay on for three states in a row.

Answer: ________________________________________

### A.3.4

Translate to LTL: Whenever the Red light is on, the Blue light will be on then or at some point in the future.

Answer: ________________________________________

### A.3.5

Translate to LTL: The Red light is on for zero or more states, and then turns off and remains off in the future.

Answer: ________________________________________

## B LTL Syntax and Semantics

Linear Temporal Logic (LTL) is a useful language for specifying the behavior of systems that change over time. An LTL formula says *what* things should happen and *when* they should happen. One run of the system satisfies a formula if and only if it does all the right things at the right times. For example, a formula for an elevator system might say that whenever the call button is pressed, the car eventually arrives at that floor. To test whether a run of the elevator system matches the formula, a formula-checker needs to inspect the sequence of elevator states and ensure that every call is followed by a matching arrival.

The LTL syntax that we use in this paper extends propositional logic with four basic temporal operators. First off, a propositional formula $P$ is one of the following:

- an atom x,
- a conjunction $P_0$ and $P_1$,
- a disjunction $P_0$ or $P_1$,
- an implication $P_0$ => $P_1$, or
- a negation not $P_0$.

The atoms come from some finite set; typically, one that describes a dynamic system. In this paper, the main set that we use consists of three atoms and describes an instrument panel with colored lights: $\{\texttt{Red}, \texttt{Green}, \texttt{Blue}\}$.

An LTL formula $Q$ is one of the following:

- a propositional formula $P$;
- an "always" formula $\mathsf{G}(Q_0)$;
- an "eventually" (or "finally") formula $\mathsf{F}(Q_0)$;
- a "next" formula $\mathsf{X}(Q_0)$;
- an "until" formula $Q_0$ U $Q_1$; or
- a conjunction, disjunction, implication, or negation of LTL formulas.

Semantically, an LTL formula describes a set of *traces*, each of which describes one run of the system at hand. We model traces as infinite sequences of sets of atoms. For example, the trace $T = T_0, T_1, T_2, \ldots$ where $T_i = \{\texttt{Red}\}$ if $i$ is even and $T_i = \{\texttt{Green}\}$ if $i$ is odd describes an execution in which the Red and Green lights alternate forever.

Let $T \gg k$ be the suffix of trace $T$ obtained by dropping the first $k$ elements. For example, $T \gg 0 = T$ and $T \gg 2 = T_2, T_3, \ldots$ (dropping $T_0$ and $T_1$). A formula $Q$ is satisfied by a trace $T$, written $T \vDash Q$, under the following conditions:

- $T \vDash P$ if $P$ holds when the atoms in $T_0$ are true and the rest are false;
- $T \vDash \mathsf{G}(Q_0)$ if $T \gg k \vDash Q_0$ for all $k \in \mathbb{N}$;
- $T \vDash \mathsf{F}(Q_0)$ if $T \gg k \vDash Q_0$ for some $k \in \mathbb{N}$;
- $T \vDash \mathsf{X}(Q_0)$ if $T \gg 1 \vDash Q_0$;
- $T \vDash Q_0 \mathsf{U} Q_1$ if $T \gg k \vDash Q_1$ for some $k \in \mathbb{N}$ and $T \gg j \vDash Q_0$ for all $j < k$; and
- likewise for conjunctions, disjunctions, implications, and negations.

This paper uses single-letter operator names and keywords interchangeably (Table 2).

## References

[1] Vicki L. Almstrum, Peter B. Henderson, Valerie J. Harvey, Cinda Heeren, William A. Marion, Charles Riedesel, Leen-Kiat Soh, and Allison Elliott Tew. "Concept Inventories in Computer Science for the Topic Discrete Mathematics". In: *ACM SIGCSE Bulletin* 38.4 (2006), pages 132–145. DOI: 10.1145/1189136.1189182.

[2] Rajeev Alur, Suguman Bansal, Osbert Bastani, and Kishor Jothimurugan. "A Framework for Transforming Specifications in Reinforcement Learning". In: *CoRR* abs/2111.00272 (2021). URL: https://arxiv.org/abs/2111.00272 (visited on 2022-09-30).

[3] Gal Amram, Suguman Bansal, Dror Fried, Lucas Martinelli Tabajara, Moshe Y. Vardi, and Gera Weiss. "Adapting Behaviors via Reactive Synthesis". In: *CAV*. Springer, 2021, pages 870–893. DOI: 10.1007/978-3-030-81685-8_41.

[4] Marco Antoniotti and Bud Mishra. "Discrete Events Models + Temporal Logic = Supervisory Controller: Automatic Synthesis of Locomotion Controllers". In: *ICRA*. IEEE, 1995, pages 1441–1446. DOI: 10.1109/ROBOT.1995.525480.

[5] Brandon Araki, Xiao Li, Kiran Vodrahalli, Jonathan A. DeCastro, Micah J. Fry, and Daniela Rus. "The Logical Options Framework". In: *ICML*. Volume 139. PMLR, 2021, pages 307–317. URL: http://proceedings.mlr.press/v139/araki21a.html.

[6] Roy Armoni, Limor Fix, Alon Flaisher, Rob Gerth, Boris Ginsburg, Tomer Kanza, Avner Landver, Sela Mador-Haim, Eli Singerman, Andreas Tiemeyer, Moshe Y. Vardi, and Yael Zbar. "The ForSpec Temporal Logic: A New Temporal Property-Specification Language". In: *TACAS*. 2002, pages 296–311. DOI: 10.1007/3-540-46002-0_21.

[7] Peter Auer, Nicolò Cesa-Bianchi, Yoav Freund, and Robert E. Schapire. "The Nonstochastic Multiarmed Bandit Problem". In: *SIAM Journal on Computing* 32.1 (2002), pages 48–77. DOI: 10.1137/S0097539701398375.

[8] Derek L. Beatty and Randal E. Bryant. "Formally Verifying a Microprocessor Using a Simulation Methodology". In: *Design Automation Conference*. ACM Press, 1994, pages 596–602. DOI: 10.1145/196244.196575.

[9] Ilan Beer, Shoham Ben-David, Cindy Eisner, Dana Fisman, Anna Gringauze, and Yoav Rodeh. "The Temporal Logic Sugar". In: *CAV*. Springer, 2001, pages 363–367. DOI: 10.1007/3-540-44585-4_33.

[10] Ilan Beer, Shoham Ben-David, Cindy Eisner, and Yoav Rodeh. "Efficient Detection of Vacuity in ACTL Formulas". In: *CAV*. Volume 1254. Springer, 1997, pages 279–290. DOI: 10.1007/3-540-63166-6_28.

[11] Amit Bhatia, Lydia E. Kavraki, and Moshe Y. Vardi. "Sampling-based motion planning with temporal goals". In: *ICRA*. IEEE, 2010, pages 2689–2696. DOI: 10.1109/ROBOT.2010.5509503.

[12] Roderick Bloem, Barbara Jobstmann, Nir Piterman, Amir Pnueli, and Yaniv Sa'ar. "Synthesis of Reactive(1) designs". In: *Journal of Computer and System Sciences* 78.3 (2012), pages 911–938. DOI: 10.1016/j.jcss.2011.08.007.

[13] Mustafa Cakir. "Constructivist Approaches to Learning in Science and Their Implications for Science Pedagogy: A Literature Review". In: *International Journal of Environmental and Science Education* 3.4 (2008), pages 193–206. URL: https://eric.ed.gov/?id=EJ894860 (visited on 2022-08-14).

[14] Hana Chockler and Ofer Strichman. "Easier and More Informative Vacuity Checks". In: *MEMOCODE*. IEEE Computer Society, 2007, pages 189–198. DOI: 10.1109/MEMCOD.2007.371225.

[15] Wonhyuk Choi, Michel Vazirani, and Mark Santolucito. "Program Synthesis for Musicians: A Usability Testbed for Temporal Logic Specifications". In: *APLAS*. Springer, 2021, pages 47–61. DOI: 10.1007/978-3-030-89051-3_4.

[16] Edmund M. Clarke and E. Allen Emerson. "Design and Synthesis of Synchronization Skeletons Using Branching-Time Temporal Logic". In: *Logics of Programs*. Volume 131. Springer, 1981, pages 52–71. DOI: 10.1007/BFb0025774.

[17] W. F. Clocksin and C. S. Mellish. *Programming in Prolog*. 4th edition. Springer–Verlag, 1984. ISBN: 978-3-540-58350-9.

[18] Jacob Cohen. "A Coefficient of Agreement for Nominal Scales". In: *Educational and Psychological Measurement* 20 (1960), pages 37–46. DOI: 10.1177/001316446002000104.

[19] Leda Cosmides and John Tooby. "Cognitive Adaptations for Social Exchange". In: *The Adapted Mind: Evolutionary Psychology and the Generation of Culture*. 1992, pages 163–228. ISBN: 978-0195101072.

[20] National Research Council. *Science Teaching Reconsidered: A Handbook*. National Academies Press, 1997. ISBN: 978-0309054980.

[21] Christoph Czepa and Uwe Zdun. "On the Understandability of Temporal Properties Formalized in Linear Temporal Logic, Property Specification Patterns and Event Processing Language". In: *IEEE Transactions on Software Engineering* 46.1 (2020), pages 100–112. DOI: 10.1109/TSE.2018.2859926.

[22] Daniel Jackson. *Alloy: a language & tool for relational models*. URL: https://alloytools.org/ (visited on 2022-08-14).

[23] Holger Danielsiek, Wolfgang Paul, and Jan Vahrenhold. "Detecting and Understanding Students' Misconceptions Related to Algorithms and Data Structures". In: *Special Interest Group on Computer Science Education*. ACM, 2012, pages 21–26. DOI: 10.1145/2157136.2157148.

[24] Richard A. DeMillo, Richard J. Lipton, and Alan J. Perlis. "Social Processes and Proofs of Theorems and Programs". In: *CACM* 22.5 (1979), pages 271–280. DOI: 10.1145/359104.359106.

[25] Paul Denny, John Hamer, Andrew Luxton-Reilly, and Helen C. Purchase. "PeerWise: Students Sharing their Multiple Choice Questions". In: *ICER*. ACM, 2008, pages 51–58. DOI: 10.1145/1404520.1404526.

[26] Matthew B. Dwyer, George S. Avrunin, and James C. Corbett. “Patterns in Property Specifications for Finite-State Verification”. In: *ICSE*. ACM, 1999, pages 411–420. DOI: 10.1145/302405.302672.

[27] Cindy Eisner and Dana Fisman. *A Practical Introduction to PSL*. Springer, 2006. ISBN: 978-0-387-35313-5.

[28] Georgios E. Fainekos, Hadas Kress-Gazit, and George J. Pappas. “Temporal Logic Motion Planning for Mobile Robots”. In: *ICRA*. IEEE, 2005, pages 2020–2025. DOI: 10.1109/ROBOT.2005.1570410.

[29] Dana Fisman, Orna Kupferman, Sarai Sheinvald-Faragy, and Moshe Y. Vardi. “A Framework for Inherent Vacuity”. In: *HVC*. Volume 5394. Springer, 2008, pages 7–22. DOI: 10.1007/978-3-642-01702-5_7.

[30] William Gasarch. *I went to the “debate” about Program Verif and the Lipton-Demillo-Perlis paper*. URL: https://blog.computationalcomplexity.org/2021/06/i-went-to-debate-about-program-verif.html (visited on 2022-03-24).

[31] Gaetano Geck, Artur Ljulin, Sebastian Peter, Jonas Schmidt, Fabian Vehlken, and Thomas Zeume. “Introduction to Iltis: an interactive, web-based system for teaching logic”. In: *ITiCSE*. ACM, 2018, pages 141–146. DOI: 10.1145/3197091.3197095.

[32] B. Glaser and A. Strauss. *The Discovery of Grounded Theory: Strategies for Qualitative Research*. Sociology Press, 1967. ISBN: 978-0202300283.

[33] Kenneth J. Goldman, Paul Gross, Cinda Heeren, Geoffrey L. Herman, Lisa C. Kaczmarczyk, Michael C. Loui, and Craig B. Zilles. “Identifying Important and Difficult Concepts in Introductory Computing Courses using a Delphi Process”. In: *Special Interest Group on Computer Science Education*. ACM, 2008, pages 256–260. DOI: 10.1145/1352135.1352226.

[34] Ben Greenman, Sam Saarinen, Tim Nelson, and Shriram Krishnamurthi. *Submitted Artifact for Little Tricky Logic: Misconceptions in the Understanding of LTL*. Version 0.2. Aug. 2022. DOI: 10.5281/zenodo.6988909.

[35] David Gundana and Hadas Kress-Gazit. “Event-Based Signal Temporal Logic Synthesis for Single and Multi-Robot Tasks”. In: *IEEE Robotics and Automation Letters* 6.2 (2021), pages 3687–3694. DOI: 10.1109/LRA.2021.3064220.

[36] Geoffrey L. Herman, Michael C. Loui, and Craig B. Zilles. “Creating the Digital Logic Concept Inventory”. In: *Special Interest Group on Computer Science Education*. ACM, 2010, pages 102–106. DOI: 10.1145/1734263.1734298.

[37] David Hestenes. “Toward a modeling theory of physics instruction”. In: *American Journal of Physics* 55.5 (1987), pages 440–454. DOI: 10.1119/1.15129.

[38] David Hestenes, Malcolm Wells, and Gregg Swackhamer. “Force concept inventory”. In: *The physics teacher* 30.3 (1992), pages 141–158. DOI: 10.1119/1.2343497.

[39] Gerard J. Holzmann. *The Spin Model Checker: Primer and Reference Manual*. Addison-Wesley, 2003. ISBN: 978-0321228628.

[40] Yatin Vasant Hoskote, Timothy Kam, Pei-Hsin Ho, and Xudong Zhao. "Coverage Estimation for Symbolic Model Checking". In: *Design Automation Conference*. ACM Press, 1999, pages 300–305. DOI: 10.1145/309847.309936.

[41] Daniel Jackson. *Software Abstractions: Logic, Language, and Analysis*. 2nd edition. MIT Press, 2012. ISBN: 978-0262528900.

[42] Yiannis Kantaros and Michael M. Zavlanos. "STyLuS$^*$: A Temporal Logic Optimal Control Synthesis Algorithm for Large-Scale Multi-Robot Systems". In: *International Journal of Robotics Research* 39.7 (2020), pages 812–836. DOI: 10.1177/0278364920913922.

[43] Juho Kim. "Improving Learning with Collective Learner Activity". PhD thesis. Massachusetts Institute of Technology, 2015. HDL: 1721.1/101464. (Visited on 2022-08-14).

[44] Juho Kim. "Organic Crowdsourcing Systems". In: *AAAI*. AAAI Press, 2016. URL: https://juhokim.com/files/AAAISymposium2016-Organic-Crowdsourcing-Systems.pdf (visited on 2022-08-14).

[45] Barbara A. Kitchenham, Shari Lawrence Pfleeger, Lesley Pickard, Peter W. Jones, David C. Hoaglin, Khaled El Emam, and Jarrett Rosenberg. "Preliminary Guidelines for Empirical Research in Software Engineering". In: *IEEE Transactions on Software Engineering* 28.8 (2002), pages 721–734. DOI: 10.1109/TSE.2002.1027796.

[46] Orna Kupferman and Moshe Y. Vardi. "Vacuity detection in temporal model checking". In: *International Journal on Software Tools for Technology Transfer* 4.2 (2003), pages 224–233. DOI: 10.1007/s100090100062.

[47] Morteza Lahijanian, Shaull Almagor, Dror Fried, Lydia Kavraki, and Moshe Vardi. "This Time the Robot Settles for a Cost: A Quantitative Approach to Temporal Logic Planning with Partial Satisfaction". In: *AAAI*. AAAI Press, 2015, pages 3664–3671. URL: https://shaull.github.io/pub/LAFKV15.pdf (visited on 2022-08-14).

[48] Leslie Lamport. "What Good is Temporal Logic?" In: *Information Processing*. North-Holland/IFIP, 1983, pages 657–668. URL: https://www.microsoft.com/en-us/research/publication/good-temporal-logic/ (visited on 2022-08-14).

[49] Savvas G. Loizou and Kostas J. Kyriakopoulos. "Automatic Synthesis of Multi-Agent Motion Tasks based on LTL Specifications". In: *CDC*. IEEE, 2004, pages 153–158. DOI: 10.1109/CDC.2004.1428622.

[50] Judith Longfield. "Discrepant Teaching Events: Using an Inquiry Stance to Address Students' Misconceptions". In: *International Journal of Teaching and Learning in Higher Education* 21.2 (2009), pages 266–271. URL: https://eric.ed.gov/?id=EJ899314 (visited on 2022-08-14).

[51] Nuno Macedo, Julien Brunel, David Chemouil, Alcino Cunha, and Denis Kuperberg. "Lightweight Specification and Analysis of Dynamic Systems with Rich Configurations". In: *FSE*. ACM, 2016, pages 373–383. DOI: 10.1145/2950290.2950318.

[52] Zohar Manna and Pierre Wolper. "Synthesis of Communicating Processes from Temporal Logic Specifications". In: *TOPLAS* 6.1 (1984), pages 68–93. DOI: 10.1145/357233.357237.

[53] Shahar Maoz and Jan Oliver Ringert. "Reactive Synthesis with Spectra: A Tutorial". In: *ICSE*. ACM, 2021, pages 320–321. DOI: 10.1109/ICSE-Companion 52605.2021.00136.

[54] Claudio Menghi, Christos Tsigkanos, Patrizio Pelliccione, Carlo Ghezzi, and Thorsten Berger. "Specification Patterns for Robotic Missions". In: *IEEE Transactions on Software Engineering* 47.10 (2021), pages 2208–2224. DOI: 10.1109/TSE.2019.2945329.

[55] Kedar S. Namjoshi. "An Efficiently Checkable, Proof-Based Formulation of Vacuity in Model Checking". In: *CAV*. Volume 3114. Springer, 2004, pages 57–69. DOI: 10.1007/978-3-540-27813-9_5.

[56] Mitchell J. Nathan, Kenneth R. Koedinger, and Martha W. Alibali. "Expert blind spot: When content knowledge eclipses pedagogical content knowledge". In: *International Conference on Cognitive Sciences*. 2001, pages 644–648. URL: http://pact.cs.cmu.edu/koedinger/pubs/2001_NathanEtAl_ICCS_EBS.pdf (visited on 2022-08-14).

[57] Mitchell J. Nathan and Anthony Petrosino. "Expert Blind Spot Among Preservice Teachers". In: *American Educational Research Journal* 40.4 (2003), pages 905–928. URL: https://www.jstor.org/stable/3699412 (visited on 2022-08-14).

[58] Joseph D. Novak and D. Bob Gowin. *Learning How to Learn*. Cambridge University Press, 1984. ISBN: 978-0521319263.

[59] Martin Oberkönig, Martin Schickel, and Hans Eveking. "A Quantitative Completeness Analysis for Property-Sets". In: *FMCAD*. IEEE Computer Society, 2007, pages 158–161. DOI: 10.1109/FAMCAD.2007.34.

[60] Liam O'Connor and Oscar Wickström. "Quickstrom: Property-based Acceptance Testing with LTL Specifications". In: *PLDI*. ACM, 2022, pages 1025–1038. DOI: 10.1145/3519939.3523728.

[61] Susan Owicki and Leslie Lamport. "Proving Liveness Properties of Concurrent Programs". In: *TOPLAS* 4.3 (1982), pages 455–495. DOI: 10.1145/357172.357178.

[62] John F. Pane and Brad A. Myers. *Usability Issues in the Design of Novice Programming Systems*. Technical report CMU-CS-96-132. Carnegie Mellon University, 1996. URL: https://apps.dtic.mil/sti/pdfs/ADA314082.pdf (visited on 2022-08-14).

[63] Roy D. Pea. "Language-Independent Conceptual "Bugs" in Novice Programming". In: *Journal of Educational Computing Research* 2.1 (1986), pages 25–36. DOI: 10.2190/689T-1R2A-X4W4-29J2.

[64] Amir Pnueli. "The Temporal Logic of Programs". In: *Foundations of Computer Science*. IEEE Computer Society, 1977, pages 46–57. DOI: 10.1109/SFCS.1977.32.

[65] Amir Pnueli and Roni Rosner. "On the Synthesis of a Reactive Module". In: *POPL*. ACM, 1989, pages 179–190. DOI: 10.1145/75277.75293.

[66] G. J. Posner, K. A. Strike, P. W. Hewson, and W. A. Gertzog. "Accommodation of a Scientific Conception: Toward a Theory of Conceptual Change". In: *Science Education* 66.2 (1982), pages 211–227. DOI: 10.1002/sce.3730660207.

[67] Akshay Rajhans, Anastasia Mavrommati, Pieter J. Mosterman, and Roberto G. Valenti. "Specification and Runtime Verification of Temporal Assessments in Simulink". In: *Runtime Verification*. Springer, 2021, pages 288–296. DOI: 10.1007/978-3-030-88494-9_17.

[68] Phyllis Reisner. "Human Factors Studies of Database Query Languages: A Survey and Assessment". In: *ACM Comput. Surv.* 13.1 (1981), pages 13–31. DOI: 10.1145/356835.356837.

[69] Sam Saarinen. "Query Strategies for Directed Graphical Models and their Application to Adaptive Testing". PhD thesis. Brown University, 2021. URL: https://repository.library.brown.edu/studio/item/bdr:kgyft3b4/ (visited on 2022-08-14).

[70] Sam Saarinen, Shriram Krishnamurthi, Kathi Fisler, and Preston Tunnell Wilson. "Harnessing the Wisdom of the Classes: Classsourcing and Machine Learning for Assessment Instrument Generation". In: *Special Interest Group on Computer Science Education*. ACM, 2019, pages 606–612. DOI: 10.1145/3287324.3287504.

[71] Leah Savion. "Clinging to Discredited Beliefs: The Larger Cognitive Story". In: *Journal of the Scholarship of Teaching and Learning* 9.1 (2009), pages 81–92. URL: https://eric.ed.gov/?id=EJ854880 (visited on 2022-08-14).

[72] Ankit Shah, Pritish Kamath, Julie A. Shah, and Shen Li. "Bayesian Inference of Temporal Task Specifications from Demonstrations". In: *NeurIPS*. 2018, pages 3808–3817. URL: https://proceedings.neurips.cc/paper/2018/hash/13168e6a2e6c84b4b7de9390c0ef5ec5-Abstract.html (visited on 2022-08-14).

[73] Abigail Siegel, Mia Santomauro, Tristan Dyer, Tim Nelson, and Shriram Krishnamurthi. "Prototyping Formal Methods Tools: A Protocol Analysis Case Study". In: *Protocols, Strands, and Logic - Essays Dedicated to Joshua Guttman on the Occasion of his 66.66th Birthday*. Springer, 2021, pages 394–413. DOI: 10.1007/978-3-030-91631-2_22.

[74] Keith Stenning and Michiel van Lambalgen. *Human Reasoning and Cognitive Science*. MIT Press, 2008. ISBN: 978-0262517591.

[75] Allison Elliott Tew and Mark Guzdial. "Developing a Validated Assessment of Fundamental CS1 Concepts". In: *Special Interest Group on Computer Science Education*. ACM, 2010, pages 97–101. DOI: 10.1145/1734263.1734297.

[76] John C. Thomas. *Quantifiers and Question-Asking*. Technical report RC-5866. IBM, 1976. URL: https://apps.dtic.mil/sti/citations/ADA043032 (visited on 2022-08-27).

[77] Moshe Y. Vardi. "An Automata-Theoretic Approach to Linear Temporal Logic". In: *Banff Workshop*. Volume 1043. Springer, 1995, pages 238–266. DOI: 10.1007/3-540-60915-6_6.

[78] Moshe Y. Vardi. “Branching vs. Linear Time: Final Showdown”. In: *TACAS*. Springer, 2001, pages 1–22. DOI: 10.1007/3-540-45319-9_1.

[79] Moshe Y. Vardi and Pierre Wolper. “An Automata-Theoretic Approach to Automatic Program Verification (Preliminary Report)”. In: *LICS*. IEEE Computer Society, 1986, pages 332–344. HDL: 2268/116609. (Visited on 2022-08-14).

[80] Moshe Y. Vardi and Pierre Wolper. “Reasoning About Infinite Computations”. In: *Information and Computation* 115.1 (1994), pages 1–37. DOI: 10.1006/inco.1994.1092.

[81] Peter Cathcart Wason. “Reasoning”. In: *New Horizons in Psychology*. Penguin, 1966. ISBN: 978-0140207750.

[82] Hillel Wayne. *Consulting*. URL: https://www.hillelwayne.com/consulting (visited on 2022-03-19).

[83] Kristin M. Weingartner and Amy M. Masnick. “Refutation Texts: Implying the Refutation of a Scientific Misconception can Facilitate Knowledge Revision”. In: *Contemporary Educational Psychology* 58 (2019), pages 138–148. DOI: 10.1016/J.CEDPSYCH.2019.03.004.

[84] Bang Wong. “Color blindness”. In: *Nature Methods* 8.6 (2011), pages 441–442. DOI: 10.1038/nmeth.1618.

[85] Tichakorn Wongpiromsarn, Alphan Ulusoy, Calin Belta, Emilio Frazzoli, and Daniela Rus. “Incremental Temporal Logic Synthesis of Control Policies for Robots Interacting with Dynamic Agents”. In: *IROS*. IEEE, 2012, pages 229–236. DOI: 10.1109/IROS.2012.6385575.

## About the authors

**Ben Greenman** (benjamin.l.greenman@gmail.com) is a postdoc at Brown University. He will be joining the University of Utah in Fall 2023.

**Sam Saarinen** (sam_saarinen@alumni.brown.edu) is a social entrepreneur whose ultimate goal is to mitigate poverty worldwide by improving the quality, equity, and scalability of education. Sam's company, Edapt Technologies, applies cutting-edge machine learning and artificial intelligence research to key educational processes, including adaptive assessment, personalized review, and individualized curriculum.

**Tim Nelson** (timothy_nelson@brown.edu) preaches the good news of logic and computing at Brown University.

**Shriram Krishnamurthi** (shriram@brown.edu) is the Vice President of Programming Languages (no, not really) at Brown University.